\documentclass[12pt]{amsart}
\usepackage[english]{babel}
\usepackage{amsmath,amssymb}
\usepackage{graphicx}
\usepackage[top=3cm,left=3cm,right=3cm,bottom=3cm]{geometry}
\usepackage{cite}

\begin{document}

\title{Multifractality and long memory of a financial index}

\author{Pablo Su\'arez-Garc\'{\i}a}
 \email{pasuarez@fis.ucm.es}
  \address{Departamento de F\'isica Te\'orica II, Universidad Complutense, 28040
Madrid, Spain.}
 \author{David G\'omez-Ullate}\thanks{This work has been partially supported by the Spanish MINECO-FEDER Grants  MTM2012-31714 and FIS2012-38949-C03-01.} 
 \email{dgullate@fis.ucm.es}
 \address{Instituto de Matem\'atica Interdisciplinar, Departamento de F\'isica Te\'orica II, Universidad Complutense, 28040
Madrid, Spain.}

\begin{abstract}

In this paper we will try to assess the multifractality displayed by the high-frequency returns of Madrid's Stock Exchange \textsc{Ibex35} index. A Multifractal Detrended Fluctuation Analysis  shows that this index has a wide singularity spectrum which is most likely caused by its long-memory. Our findings also show that this long-memory can be considered as the superposition of a high-frequency component\,---related to the daily cycles of arrival of information to the market---\,over a slowly-varying component that reverberates for long periods of  time and which shows no apparent relation with human/economic cycles. This later component is therefore postulated to be endogenous to market's dynamics and to be also the most probable source of some of the stylized facts commonly associated with financial time-series.\\[8pt]


\end{abstract}
\maketitle

\section{Introduction}
\label{Introduction}

Many of the time-series observed in a wide range of natural and social phenomena have been described as \emph{intermittent}, a concept coined in the field of fluid mechanics where this behavior was first reported \cite{batchelor1953the}. Intermittent signals display periods of high and violent activity interspersed with periods where events are sparser and smaller in size, with this alternation depending on the scale under consideration.
 Therefore, small and big fluctuations exhibit different scaling behaviors and so their relative frequency changes along with the scale.  This erratically changing oscillatory behaviour and heterogenity in the level of fluctuation  is in stark contrast with the usual self-similar (\emph{fractal}) behavior exhibited by ``normal'' or gaussian time-series, where the fluctuations at fine scales are uniform and  all of them follow the same scaling law.

The \emph{multifractal} model is one of the many proposed to explain the intermittency observed in fully developed turbulence \cite{mandelbrot1974intermittent, parisi1985turbulence, frisch1995turbulence} although it has also found applications in a wide range of fields: geophysics and meteorology \cite{mandelbrot1989multifractal, tessier1996multifractal,  singh2009experimental},  \textsc{mhd} fluctuations of the solar wind \cite{marsch1997intermittency}, heartbeat dynamics \cite{stanley1999statistical}, \textsc{dna}  sequences \cite{arneodo1998thermodynamics}, network traffic \cite{gilbert1999scaling}, image modeling \cite{turiel2000multifractal} and natural language \cite{ausloos2012measuring} \emph{inter alia}.  As an opposition to \emph{fractal} processes where all  fluctuations follow the same scaling law and so they can be described with a single \emph{scaling exponent}, the term \emph{multifractal} captures the manifold, location dependent and erratically changing local  scaling exponents\,---which are, in turn,  distributed in a scale-invariant way---\,displayed by intermittent signals. The \emph{multifractal formalism} is used to study the local (resp. global) scaling properties of these time-series in geometrical terms via its local scaling exponents (resp.  in statistical terms via  its sample moments)  \cite{riedi2010multifractal}. It is worth mentioning that this formalism also unfolds a not-so-obvious connection between these two diametrically opposed points of view.

It is also well documented that financial time-series exhibit intermittence as well as other non-trivial statistical properties collectivelly known as \emph{stylized facts} \cite{cont2001empirical}. Some of the most celebrated  are the  \emph{leptokurtic probability distributions}, their \emph{scaling symmetry} and \emph{long term agregational gaussianity}, the \emph{martingale property}, the \emph{clustering of volatilities} and the \emph{long memory of the nonlinear moments}, to name just a few. In a series of seminal papers \cite{calvet1997large, fisher1997multifractality, mandelbrot1997multifractal}, Calvet, Fisher and Mandelbrot laid the foundations (both theoretical and practical) of multifractal analysis applied to financial time-series. They also proposed the \emph{multifractal model of asset returns}, that claims to explain most of the stylized facts observed in financial time-series.  In short, the model considers  that the logarithm of a financial price reduces to a (possibly fractional) Brownian motion $B_H(\theta)$ (being $H$ the \emph{Hurst exponent} of the fractional Brownian motion) in terms of a \emph{trading time} $\theta(t)$, which is in turn a multifractal non-decreasing function of the clock time $t$.  Since then, their ideas have been thoroughly tested and corroborated by many studies \cite{ lux2003multi, xu2003scaling, wang2012statistical, norouzzadeh2006multifractal, zhou2009components, matia2007multifractal, turiel2003multifractal, wang2012multifractal,eisler2004multifractal}.

In this paper we are going to assess the multifractality displayed by a financial time-series and we will try to pinpoint its origin.  It is organized as follows:  after reviewing the basic ideas behind the multifractal formalism in the second section, we will analyze the local singular behavior of the time-series to test for multifractality and we will obtain its scaling exponents and singularity spectrum (Section \ref{sect:Results}). In this section we will also address the origins of this multifractality. In Section \ref{sect:Discusion} we will sum up the results of the previous section and we will confront them with those obtained in other studies. The conclusions will be expounded in Section \ref{sect:Conclusions}.


\section{The Multifractal Formalism} \label{sect:multifractalformalism}

\subsection{H\"older exponents and the singularity spectrum} \label{subsect:holder}
The local roughness of a function $X(t)$ or a time-series is given by its \emph{H\"older} or \emph{singularity exponent}, which is defined at each point $t$ by the exponent $\alpha(t)$ such that for $\epsilon >0$: 
\begin{equation}
|X(t+\epsilon)-X(t)| \sim C(t)\epsilon^{\alpha(t)}, \hspace{5mm}  \epsilon \rightarrow 0
\end{equation}
The  H\"older exponent  at point $t$  describes the local  scaling properties of the function  $X(t)$ and measures its smoothness or degree of differentiability. When $\alpha(t)=1$ the function is differentiable at $t$; when $\alpha(t)=0$, the function is discontinuous at that point. Values of this exponent  between $0$ and $1$ indicate that the function is continuous but not differentiable, being  smoother at points $t$ where $\alpha(t)$ approaches $1$ and rougher if $\alpha(t)$ goes to $0$. Any (mono-)fractal object (\emph{e.g.} Brownian motion paths)  has the same exponent for all $t$ (\emph{e.g.} $\alpha(t)=\frac{1}{2}$ for a Brownian motion), a fact that indicates that its fluctuations are uniform and scale-independent. 

A multifractal, however, cannot be described with a single singularity exponent; a continuum of them is needed to characterize its scaling behavior. In a multifractal object (\emph{viz.} a set, a function, a time-series, etc.), the set of points  with a given singularity exponent $\alpha$ is a fractal with fractal dimension $f(\alpha)$. These sets are knitted together yielding a myriad of intertwined fractals\,---hence the term \emph{multifractal}---\,making fluctuations non-uniform and its relative frequency scale-dependent. With the  adequate renormalization\footnote{Obviously, all the subsets with fractal dimension $f(\alpha)<1$ have measure zero and therefore its measure cannot be directly used as a proxy for its relative frequency.},  $f(\alpha)$\, ---the singularity spectrum---\, describes the relative occurrence of local singularities with exponent $\alpha$; the wider it is, the more heterogeneous the fluctuations.  It is in this sense how the spectrum of singularities encodes the average roughness and the intermittency of the signal. 


\subsection{Global scaling analysis and the partition function}

Unfortunately, direct calculation of   $f(\alpha)$ is not easy and it is prone to numerical errors \cite{riedi1999introduction}. A different route has to be taken to solve this problem that, as a bonus,  reveals a useful relation between  $f(\alpha)$ and the scaling behaviour of the statistical moments of the process \cite{riedi2010multifractal}. In fact, multifractal stochastic processes are usually defined \cite{mandelbrot1997multifractal, calvet1997large, calvet2002multifractality} as those processes with stationary increments whose moments satisfy the following scaling law:

\begin{equation}
\mathbf{E}[|X(t)|^q]=c(q)t^{\tau(q)+1} \hspace{5mm} t, q \in \mathbf{R} \label{eq:multifractal}
\end{equation}

A (mono-)fractal object has a scaling exponent $\tau(q)$ that depends linearly on $q$, namely, $\tau(q)=qH$. For a multifractal object, however, this relation is nonlinear. To obtain the exponents $\tau(q)$ and relate them to the multifractal spectrum a method called the \emph{partition function} is generally used. 

For a given function or time-series $X(t)$ defined for $t \in [0,T]$ and for each scale $\epsilon>0$ we define  the fluctuation\footnote{In practice, instead of the fluctuations  $S_{X}(t;\epsilon)=X(t+\epsilon)-X(t)$ other \emph{generalized fluctuations}  are often preferred. Examples of methods for scaling analysis that use these generalized fluctuations are the \emph{Wavelet Transform Modulus Maxima} \cite{muzy1993multifractal}, where the continuous wavelet transform of the function takes the role of the fluctuations, or the \emph{Multifractal Detrended Fluctuation Analysis} \cite{kantelhardt2002multifractal}, which will be developed later on. Both of them are more robust against small-scale noise or long-range trends than the plain fluctuations.}  of the process at time $t$: 
\begin{equation}
S_{X}(t;\epsilon)=X(t+\epsilon)-X(t)
\end{equation}

Next we partition $[0,T]$ in $N$ intervals of size $\epsilon$ and we define the \emph{sample sum} or \emph{partition function} of $q^{th}$ order:

\begin{equation}
Z_q(\epsilon)=\sum_{n=0}^{N-1}|S_{X}(n\epsilon;\epsilon)|^{q} \label{eq:partition}
\end{equation}

Now, if $X(t)$ is a multifractal process, it is stationary and  the following holds (\emph{cf.} equation \ref{eq:multifractal}):

\begin{equation}
\mathbf{E}[Z_q(\epsilon)]=Tc(q)\epsilon^{\tau(q)} \label{eq:partitionscaling}
\end{equation}

The exponent $\tau(q)$ can then be estimated from the slope of the log-log plot of  $Z_q(\epsilon)$ against the box size $\epsilon$ via ordinary least-squares regression. Nevertheless, the consistency of this method of inference is still under debate \cite{calvet1997large}. For other generalized fluctuations (\emph{e.g.} \textsc{mf-dfa} or Wavelet Transform Modulus Maxima) the exponents obtained by this scaling analysis are algebraically related to the $\tau(q)$.


\subsection{Relationship between the two descriptions}

From the exponents  $\tau(q)$ the singularity spectrum  $(\alpha, f(\alpha))$  can be recovered by the following contact transformation   \cite{calvet1997large}:

\begin{equation}
\left.\frac{df(\alpha)}{d\alpha}\right| _{\alpha_q}=q, \hspace{5mm}  \tau(q)=q\alpha_q - f(\alpha_q) \label{eq:legendre}
\end{equation}

which is nothing but the Legendre transform.  This shows that the singularity spectrum is the envelope of the family of straight lines with slope $q$ and $y$-intercept $-\tau(q)$, with $\alpha_q$ being the abscissa of their corresponding points on the envelope. The value $\alpha_0$ verifies $f(\alpha_0)=1$ (\emph{cf.} Equation (\ref{eq:multifractal})), and so it  is the H\"older exponent with non-zero measure, the ``most typical'' exponent.

This unexpected connection between the geometry  of the set (the \emph{local} roughness of the signal as described by its singularity spectrum)  and its statistics (the \emph{global} scaling of the sample moments $\tau(q)$)  is known as \emph{the multifractal formalism} \cite{riedi1999multifractal}. It is a consequence of the  \emph{large deviations principle}  \cite{calvet1997large}, although the intuition behind it is relatively straightforward: only the points with exponent $\alpha_q$  contribute to the sum (\ref{eq:partition}) for the $q^{th}$-order moment \cite{tel1988fractals}.


\subsection{Multifractal Detrended Fluctuation Analysis} \label{subsect:MFDFA}
One of the most widely used methods to estimate the singularity spectrum is the \emph{Multifractal Detrended Fluctuation Analysis} (\textsc{mf-dfa}), which is robust against non-stationarities or trends in the series. It was first developed in  \cite{kantelhardt2002multifractal}, and it is based on the more usual \emph{Detrended Fluctuation Analysis} \cite{kantelhardt2001detecting}. For the sake of completeness, we shall review it shortly. 

First, the time series is divided in $N_s$ non-overlapping segments of length $s$. Next, a polynomial of order $n$ is fitted by least squares to each of these segments. The $q^{th}$ order fluctuation $F_q(s)$ function is the $q^{th}$-power mean of $F(s)$, the root mean square of the residuals  of the regression over each of the $N_s$ segments.  For a multifractal process, $F_q(s)$ scales as:

\begin{equation}
F_q(s) \sim s^{h(q)}
\end{equation}

The scaling exponents $h(q)$ obtained by this method are directly related to the $\tau(q)$ scaling exponents defined by the standard partition function-based formalism by the following transformation:
\begin{equation}
\tau(q)=qh(q)-1
\end{equation}

The singularity spectrum can now be obtained  via the Legendre transform.


\section{Analysis of the data} \label{sect:Results}

\subsection{The data}

The time series analyzed in this study has been already used elsewhere \cite{suarez2013scaling} to study its distributional properties and scaling symmetries. There it was shown that its distribution function displayed a strong non-gaussianity and  scaling symmetry over a wide range of scales despite having non-stable power laws.

The data set contains the price ticks of the index \textsc{Ibex35} of the Madrid Stock Exchange\footnote{Data obtained from \emph{www.tickdata.com}}. The index \textsc{Ibex35} is a weighted index formed by the 35 most liquid Spanish stocks traded at the Madrid Stock Exchange. The data set covers the period from January 2nd 2009 to December 31st 2010 and comprises 510 market days. In order to have a well defined time interval we have sampled these ticks in fifteen-seconds intervals obtaining a series with 1036321 records. From this series we have obtained the log-returns, focusing only on intra-day returns and therefore discarding the discontinuity created overnight by the closing of the market and taking into account the 30-second uncertainty in the closing time of the session in order to avoid arbitrages.  
 
 Even though the series displays strong non-stationarities related to the periodic arrival of information to the market (\emph{lunch effect}), we have decided to work with the raw returns instead of trying to detrend them. The rationale behind this decision is two-fold. First, it is precisely this seemingly non-stationary behaviour one of the stylized facts that a multifractal time-series  explains without resorting to a hypothesis of non-stationarity. And second, in case there existed real trends in the data, the method used to study its multifractality (\textsc{mf-dfa}) would factor them out. The final return series thus obtained contains 1035810 records, with 2031 records for each market day.
 

\subsection{Estimation of the singularity spectrum}

\begin{figure}[h!]
    \includegraphics[width=\textwidth]{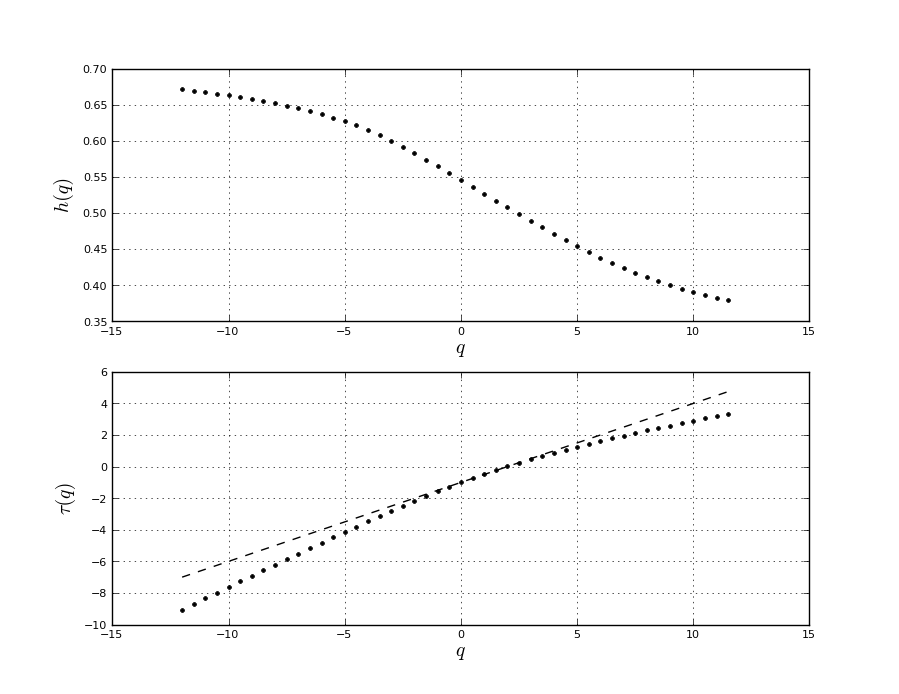}\caption{Top panel: $h(q)$. Bottom panel: $\tau (q)$. The dashed line is the expected $\tau(q)$ for a process with stationary and independent increments with finite variance.}
 \label{figure:taus}
\end{figure}

\begin{figure}[h!]
    \includegraphics[width=\textwidth]{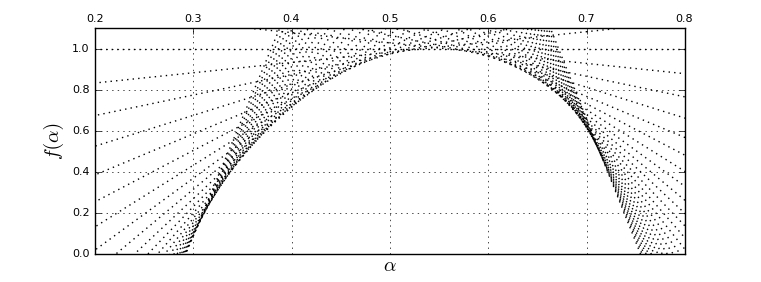}\caption{Singularity spectrum $(\alpha, f(\alpha))$ as the envelope of the lines $y(\alpha) = q\alpha-\tau(q)$.}
 \label{figure:spectrum}
\end{figure}

To estimate the singularity spectrum we have used the method of \emph{Multifractal Detrended Fluctuation Analysis} (\textsc{mf-dfa})  (\emph{cf.} section \ref{subsect:MFDFA}). For this study, we have chosen $5$ as the order of the fitting polynomial. The functions $h(q)$ and $\tau(q)$ are shown in figure \ref{figure:taus}.

From the figure we readily see that the function $\tau(q)$ is clearly nonlinear, an indicator of the multifractality displayed by the time-series. By means of the Legendre transform of $(q, \tau(q))$ we can obtain the singularity spectrum. Figure \ref{figure:spectrum} displays the results. The numerical properties of the spectrum are summarized in Table \ref{table:spectrum}. The width of the spectrum is a clear evidence of the multifractality of the signal.

\begin{table}
\renewcommand{\arraystretch}{1.2}
\centering
\begin{tabular}{ccc}
\hline
{ $\mathbf{\alpha^{min}}$} & {$\mathbf{\alpha^{MAX}}$} & {$\mathbf{\alpha_{0}}$}  \\ 
\hline 
0.300 & 0.717 & 0.545   \\
\hline
\end{tabular}
\vspace{5pt}
\caption{\textbf{Spectrum properties.}}
\label{table:spectrum} 
\end{table}


\subsection{Sources of multifractality}

As it is widely acknowledged \cite{mandelbrot2001scalingIII, kantelhardt2002multifractal}, there are two different sources of multifractality in a time-series. It can be \emph{(a)} due to a fat-tailed probability distribution or \emph{(b)} caused by the different long-range correlations of the small and large fluctuations. The best way to distinguish one source from the other is by generating surrogate data sets  \cite{theiler1992testing}. Reshuffling the series destroys the time correlations although preserves the distribution of increments: therefore, it is a proxy of the contribution of the fat-tailedness of the distribution to its multifractality.  As it was shown in  \cite{suarez2013scaling}, it can be assumed  that the tails of the distribution of log-returns are far from the stability regime, even despite all the pitfalls endemic  to the estimation of this parameter \cite{dumouchel1983estimating, mcculloch1997measuring, malevergne2005empirical}.

In any case, and to get more evidence in favor of this hypothesis, by reshuffling the original data we have produced a collection of $10^4$ surrogate data-sets with the same distribution as our time-series but without its time correlations. With this, we have obtained the empirical distribution of $ d = \sqrt{\tau(q)-(\frac{q}{2}-1)}$,  our discriminating statistic  (\emph{cf.}  Table \ref{table:surrogate}). This is simply the $l^2$ distance of $\tau(q)$ to  $(\frac{q}{2}-1)$,  the expected $\tau(q)$ for a null hypothesis of i.i.d. random variables with finite variance. The statistic  $ d$ is a measure of the multifractality of the data-set.  The observed value for our time-series ($\hat d  = 3.221$) yields a $p$-value less than  $10^{-4}$, so it is safe to assume that the observed mulifractality cannot be explained by the fat-tailedness.

\begin{table}
\renewcommand{\arraystretch}{1.2}
\centering
\begin{tabular}{cccccc}
\hline
{ $\mathbf{\mu}$} & {$\mathbf{\sigma}$} & {$\mathbf{min.}$} & {$\mathbf{MAX.}$} & {$p_{99}$} &  {$p_{99.9}$}  \\ 
\hline 
0.523 & 0.302 & 0.009 & 2.174 & 1.310 & 1.590 \\
\hline
\end{tabular}
\vspace{5pt}

\caption{\textbf{Distribution of  $d$.} Summary statistics and $p_i$ percentiles.}
\label{table:surrogate} 
\end{table}

\begin{figure}[h!]
    \includegraphics[width=\textwidth]{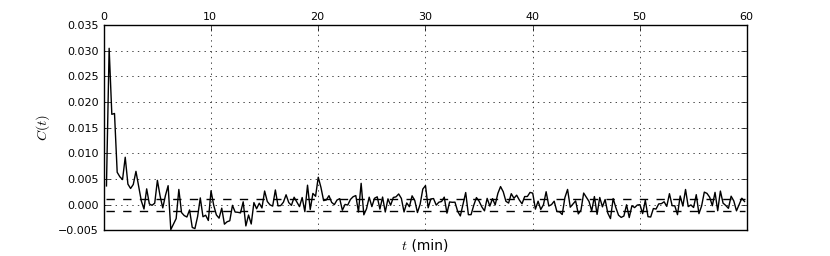}\caption{Correlogram of \textsc{Ibex35} raw returns. The dashed lines are the $95\%$ confidence interval for white noise.} \label{figure:correlogram}
\end{figure}

\begin{figure}[h!]
    \includegraphics[width=\textwidth]{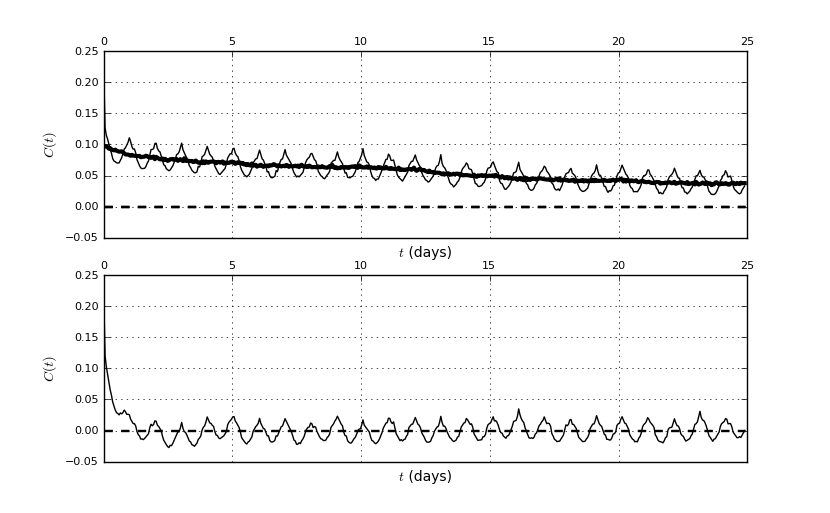}\caption{Correlogram of absolute returns. Top: absolute returns and absolute returns intra-daily reshuffled (in bold). Bottom: absolute returns daily reshuffled. The dashed lines are the $95\%$ confidence interval for white noise.}
 \label{figure:abscorrelogram}
\end{figure}

Given this, we are left with the temporal correlations of the time-series as the most likely responsible of the multifractality observed in it. As it happens with many other financial time-series \cite{cont2001empirical}, the \textsc{Ibex35} index can be considered as serially uncorrelated  (\emph{cf.} Figure \ref{figure:correlogram}). The nonlinear moments, however, are an entirely different issue: the correlations of \emph{e.g.} the absolute returns display a long memory that reverberates over long periods of time  (\emph{cf.} Figure \ref{figure:abscorrelogram}) and after 25 days those correlations are nowhere close to the $95\%$ confidence interval for white noise. 

\begin{figure}[h!]
\centering
    \includegraphics[width=\textwidth]{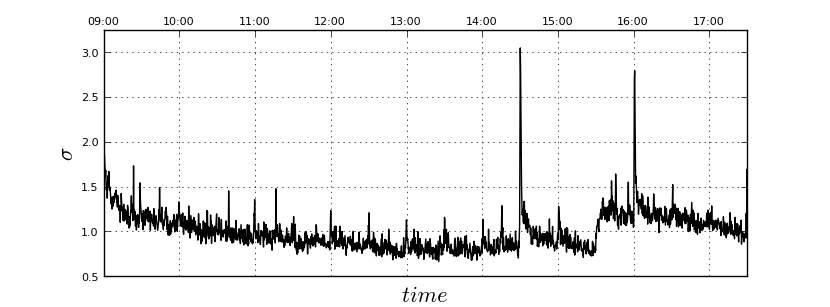}\caption{\textbf{Lunch effect.} At 15:30 CET Wall Street opens, and at 14:30 CET and 16:00 CET macroeconomic indicators in the USA are announced.}
 \label{figure:lunch}

\end{figure}

Part of this long memory is caused by the intraday average-volatility patterns present in the data related with the arrival of information to the market and its activity (\emph{i.e.} the \emph{lunch effect}): the market opens at 09:00 CET and closes at 17:30 CET and so it is in sync with human cycles. In the meantime, economic news arrive at precise moments during the session and also different forward contracts reach maturity at established times  (\emph{cf.} Figure \ref{figure:lunch}).  This cyclical pattern has an obvious influence over the long memory displayed by the index. However, it is neither its sole nor its principal cause: the figures clearly  show that reshuffling the data intra-daily destroys this patterns but the most significant correlations still remain.

\begin{figure}[h!]
    \includegraphics[width=\textwidth]{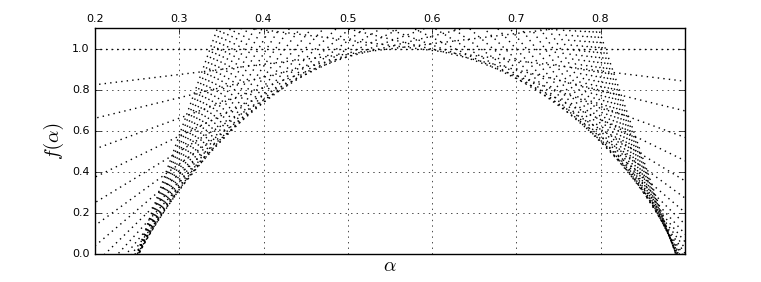}\caption{Singularity spectrum $(\alpha, f(\alpha))$ for intra-daily reshuffled returns.}
 \label{figure:spectrum_shuff}
\end{figure}

In the light of these figures, the correlation of absolute returns could be considered as composed of a high-frequency component caused by the intra-daily patterns plus a slowly-varying component spreading over many scales. This component is in fact the one that carries most of the multifractality. Indeed, the multifractality is even more dramatic when the intra-daily patterns are factored-out (\emph{cf.} Figure \ref{figure:spectrum_shuff}). As it was shown in  \cite{suarez2013scaling}, this operation did not affect  the scaling symmetry displayed by the distribution of log-returns either. However, reshuffling the whole dataset destroyed the index's scaling symmetry for good.


\section{Discussion}
\label{sect:Discusion}

The multifractality of financial time-series has already been assessed for many assets, markets or time-periods: FX markets \cite{mandelbrot1997multifractal, lux2003multi, xu2003scaling, wang2012statistical, norouzzadeh2006multifractal}, stock-market indexes \cite{ausloos2002multifractal, lux2003multi, zhou2009components, caraiani2012evidence}, commodities like gold or electricity \cite{lux2003multi, norouzzadeh2007anti}, single stocks \cite{turiel2003multifractal} or growth enterprise markets \cite{wang2012multifractal} \emph{inter alia}. In what regards  the \textsc{Ibex35} index,  we have found significant deviations of $\tau(q)$ from $q/2-1$ and and a wide singularity spectrum $f(\alpha)$. These  are both quite strong evidences in  favor of its multifractality.

In a few of the above cited papers, the origin of the multifractality displayed by the series has been also addressed. However, consensus about this issue has not been reached yet. While in our study the multifractality of the  time-series is most likely to be due to its temporal correlations, in the daily returns of the \textsc{Dow Jones} it seems  to be caused by the fat-tailedness of the distribution of returns \cite{zhou2009components}. Other studies are not as conclusive: in  \cite{matia2007multifractal} the multifractal spectra of stocks and commodities is partly related to the power law probability distribution of returns and partly to the higher order temporal correlations present on it, and in  \cite{norouzzadeh2006multifractal}, although most of the multifractality encountered is due to different long-range correlations for small and large fluctuations, the shape of the probability distribution function also contributes to the multifractal behaviour of the time series. Finally, \cite{wang2012statistical} showed that a legislation change can alter the equilibrium between the contributions of the fat-tailedness of the distribution of returns and its long memory. As it can be seen, the question remains a moot point.

A closely related issue to the one just expounded is the influence that the arrival of information has on market dynamics.   It is quite well known that volatility is much too high to be explained by changes in fundamentals  \cite{shiller1980stock} and that most of price movements  cannot be explained only with news \cite{cutler1998moves}. What we have found in this study is that, although the arrival of information clearly affects the volatility during the session and is thus also responsible for the striking daily patterns found in the autocorrelation function of the absolute returns, there is nevertheless another component on this autocorrelation function. This slowly-varying component  does not seem to be related to human/economic/information cycles at all; notwithstanding, it is the responsible of most of the multifractality observed on the time-series and it seems to be also deeply related to other stylized facts observed in it like its scaling symmetry. It has been proposed  \cite{bouchaud2010endogenous} that the erratic dynamics of markets is to a large extent of endogenous origin, \emph{i.e.} determined by the trading activity itself and not due to the rational processing of exogenous news, inline with our findings. Part (or all) of this effect might be caused by the way traders interact, leading to notions like the ``reflexivity'' of investorsÕ decisions \cite{soros2003alchemy}.


\section{Summary}
\label{sect:Conclusions}

We have performed a multifractal and correlation analysis of the high-frequency returns of the \textsc{Ibex35} index of Madrid stock exchange over a two year period (2009-2010). By means of a Multifractal Detrended Fluctuation Analysis, the scaling exponents $\tau(q)$ and its singularity spectrum $f(\alpha)$ have been obtained. Both show that the time-series displays a clear multifractality.

The analysis of linear correlations of the index shows that it can be considered as serially uncorrelated, as usually happens with many other financial time-series. The nonlinear correlations are however quite a different issue: they display long-memory over very long periods, even as long as the time period analyzed. Apart from that, the autocorrelation function of absolute returns displays a striking daily pattern caused by the cyclical nature the stock market session (the \emph{lunch effect}) superimposed on a slowly varying component. Furthermore, this component seems to cause most of the multifractality and  does not seem to be related to human/economic/information cycles at all. It is therefore postulated to be endogenous to market's dynamics and not related to exogenous triggers.

\section{Acknowledgements}

The research of DGU has been
supported in part by Spanish MINECO-FEDER Grants  MTM2012-31714 and FIS2012-38949-C03-01


\end{document}